\documentclass[aps,prd,nofootinbib,twocolumn,floatfix,eqsecnum]{revtex4}
\usepackage{dcolumn}
\usepackage{times,mathptm}
\usepackage{subfigure}
\usepackage{graphicx,psfrag,pifont,color,setspace}
\usepackage{amsfonts,amsthm,bm,bbm,amsmath,amssymb,marvosym} 
\usepackage[automark]{scrpage2}
\newcommand{\beq}{\begin{equation}}
\newcommand{\eeq}{\end{equation}}
\newcommand{\bea}{\begin{eqnarray}}
\newcommand{\eea}{\end{eqnarray}}
\newtheorem*{theorem}{Theorem}

\renewcommand{\d}{\delta}
\renewcommand{\l}{\lambda}

\renewcommand{\b}{\beta}
\renewcommand{\a}{\alpha}

\renewcommand{\ni}{\noindent}

\newcommand{\m}{\mu}
\renewcommand{\r}{\rho}

\newcommand{\bx}{{\mathbf{x}}}
\newcommand{\by}{{\mathbf{y}}}

\newcommand{\bn}{{\mathbf{n}}}
\newcommand{\bp}{{\mathbf{p}}}
\newcommand{\hi}{\hat{i}}

\newcommand{\s}{\sigma}
\newcommand{\A}{{\cal A}}
\newcommand{\D}{{\Delta}}

\newcommand{\tD}{\widetilde{D}}
\newcommand{\tA}{\widetilde{A}}

\renewcommand{\th}{\theta}

\newcommand{\e}{\epsilon}

\newcommand{\oh}{{\textstyle{\frac{1}{2}}}}

\newcommand{\oq}{{\textstyle{\frac{1}{4}}}}

\newcommand{\dg}{\dagger}
\newcommand{\non}{\nonumber}

\newcommand{\rf}[1]{(\ref{#1})}
\newcommand{\ra}{\rightarrow}
\newcommand{\pa}{\partial}

\begin{document}

\title{Faddeev-Popov spectra at the Gribov horizon}

\author{J. Greensite}
\affiliation{Physics and Astronomy Dept., San Francisco State
University, San Francisco, CA~94132, USA}

%
\date{\today}
\begin{abstract}
   I present a perturbative calculation of the spectrum of the Faddeev-Popov
operator in Coulomb gauge in three dimensions, and Landau gauge in two and 
three dimensions, with an ansatz for the gluon propagator as the  non-perturbative input. 
The results show how the low-lying Faddeev-Popov eigenvalue spectrum is modified as 
the first Gribov horizon is approached, and how the spectra can differ in Coulomb and Landau 
gauges.
\end{abstract}

\maketitle

\section{\label{intro}Introduction}

        The low-lying spectrum of the Faddeev-Popov (F-P) operator, in Coulomb and covariant gauges, is a probe of the infrared
properties of non-abelian gauge theories.  Confinement in Coulomb gauge, in particular, is rather directly related to the F-P
spectrum. The color Coulomb potential, for example, involves a product of inverse F-P operators, and the Coulombic self-energy 
of an isolated color charge, which is infrared divergent in a confining theory, depends crucially on the density of low-lying eigenvalues 
of the F-P operator, as discussed below.  The connection to confinement is less apparent in covariant gauges, although the density 
of near-zero F-P eigenmodes is expected to be relevant to the infrared behavior of the ghost propagator.

   Coulomb and Laudau gauges are defined on the lattice as the set of elements, on each gauge orbit, for which the quantity
\beq
           R[U] = - \sum_{x} \sum_{\m=1}^d \text{Tr}[U_\m(x)]
\label{R}
\eeq
is stationary with respect to infinitesimal gauge transformations.  Here $d$ denotes the number of space dimensions, in Coulomb gauge, and the number of spacetime dimensions in Landau gauge, which is a convention I will adopt from now on.  In general, along any gauge orbit, there are many stationary points, known as Gribov copies, and at these points the F-P determinant may be positive or negative.  This indefinite sign is closely to related to Neuberger's Theorem \cite{Herbert}, which demonstrates that BRST quantization of any non-abelian lattice gauge theory  is ill-defined at the non-perturbative level.  The picture is that in summing over all copies on a gauge orbit, the copies with a positive F-P determinant are exactly cancelled by the copies with a negative F-P determinant, and the functional integral vanishes.   It is for this reason that a constraint of some kind, imposed on the domain of functional integration, is necessary.   Ideally the range of functional integration should be a subspace (such as the Fundamental Modular region) containing only a single gauge copy with positive F-P determinant per gauge orbit, but at a minimum the integration range should lie within the Gribov region.  This is the region which consists of all gauge copies in which the non-trivial eigenvalues of the F-P operator are all positive; i.e.\ the Gribov copies which are local minima of $R[U]$.  These are, in fact, the configurations obtained by standard lattice gauge-fixing algorithms.  The Gribov region is completely bounded, and the Fundamental Modular region is partially bounded, by the first Gribov horizon, where the lowest non-trivial F-P eigenvalue vanishes.  It has been argued by Zwanziger \cite{Dan1} that the volume of the Gribov region is concentrated close to the horizon, much as the volume of a sphere in a high dimensional Euclidean space is concentrated near the surface.  Since the dimensionality of the space of all lattice configurations is very high indeed, the values of observables obtained at the Gribov horizon should dominate the expectation value.   It would then be interesting to understand exactly how proximity to the Gribov horizon affects the behavior of various observables, starting with the spectrum of the F-P operator.
    
    As a step in that direction, this article presents a perturbative calculation of the F-P spectra.  Perturbation theory is not necessarily trustworthy when dealing with the low-lying eigenmodes, but something may still be learned from it.  In particular, it would be interesting to see whether proximity to the Gribov horizon changes the behavior of the low-lying spectra already at the perturbative level, and whether that behavior is different, for some reason, in Coulomb and Landau gauges.  The calculation is carried out for Landau gauge in two and three spacetime dimensions, and Coulomb gauge in three spacetime dimensions, to avoid the complications associated with renormalization in four dimensions.\footnote{Yang-Mills theory in Coulomb gauge is trivial in two spacetime dimensions if the spacetime manifold is flat and non-compact, and for that reason Coulomb gauge in $D=2$ will not be considered here.  Non-trivial features associated with the Coulomb gauge F-P operator do appear even in two dimensions, if the space direction is compactified to $S^1$, and this case has been thoroughly discussed by Reinhardt and Schleifenbaum in ref.\ \cite{oneone}.} The proximity to the Gribov horizon is controlled by a mass parameter in the transverse gluon propagator, which is where the non-perturbative information enters.  I use an ansatz for the gluon propagator, motivated by Gribov's expression \cite{Gribov}, which allows for any desired power behavior in the infrared.   

\section{\label{FPC} F-P eigenvalues and the Coulomb self-energy}

The Coulomb potential between a static quark-antiquark pair located at points $x$ and $y$ is given by the expression
\beq
          V_C(|x-y|)  = -{g^2 C_r \over d_A}   \left\langle (M^{-1})^{ab}_{xz} (-\nabla_z^2) (M^{-1})^{ba}_{zy} \right\rangle
\eeq
where $C_r$ is the quadratic Casimir of quarks in color representation $r$, $d_A$ is the dimension of the adjoint representation of
the gauge group, and $M$ is the Faddeev-Popov operator, which is
\beq
         M^{ab}_{xy} = \Bigl(-\d^{ab} \nabla^2 + gf^{abc} A^c_i(x) \pa_i \Bigr) \d^3(x-y)
\label{FPcont}
\eeq
in the continuum.  If $V_C(|x-y|)$ is confining, then this can only be attributed to an infrared singular behavior of $M^{-1}$, which must be related
somehow to the low-lying F-P eigenvalue spectrum.  

        The perturbative evaluation of the F-P spectrum starts with the free-field, $g^2=0$ case, on a finite periodic lattice of extension $L$. 
The eigenmodes of the corresponding F-P operator are simply the plane wave states     
\bea
         \phi_{\bn,A}^{a(0)} &=& {1\over \sqrt V} e^{i p \cdot x } \chi^a_A 
\non \\
          \l_{\bn,A}^{(0)} &=& 2\sum_\m (1-\cos(p_\m))
\non \\
          p_i &=& 2\pi {n_i \over L} ~~,~~ -{L\over 2} < n_i \le  {L\over 2}
\label{zeroth}
\eea 
and the $\vec{\chi}_A$ are some set of orthonormal vectors spanning the $d_A$-dimensional color space.  The F-P eigenmodes and
eigenvalues $\{\phi_{\bn,A}(x),\l_{\bn,A}\}$ at $g^2 >0$ are also indexed by $(\bn,A)$, denoted for brevity by $n \equiv (\bn,A)$, and it
is assumed that the eigenmodes and eigenvalues are continuous and differentiable functions of $g^2$, which smoothly approach \rf{zeroth}
as $g^2\ra 0$.   To connect the eigenmode spectrum to the Coulomb self-energy, we begin with the expression
\beq
          E_{self} = {g^2 C_r \over d_A}  \lim_{V\ra \infty} {1\over V} \left\langle (M^{-1})^{ab}_{xz} (-\nabla_z^2) (M^{-1})^{ba}_{zx} \right\rangle
\eeq
and inserting the spectral representation
\beq
           (M^{-1})^{ab}_{xy} = \sum_n {\phi^a_n(x) \phi^{*b}_n(y) \over \l_n}
\eeq
this becomes \cite{GOZ2}
\bea
          E_{self} &=&  {g^2 C_r \over d_A} \lim_{V\ra \infty}  {1\over N_c  V} \sum_n  \left\langle {(\phi_n|-\nabla^2|\phi_n) \over \l_n} \right\rangle
\non \\
&=& g^2 {C_r\over d_A} \int_0^{\l_{max}} d \l ~\left\langle \rho(\l) {(\phi_\l|-\nabla^2|\phi_\l) \over \l^2} \right\rangle
\non \\
\label{Es}
\eea
where $\rho(\l)$ is the normalized eigenvalue density
\beq
           \rho(\l) = \lim_{V\ra \infty}  {1\over N_c V} \sum_n \d(\l - \l_n)
\eeq 
$N_c$ is the number of colors, and $V=L^d$.
In 2+1 dimensions, the integral in \rf{Es} is logarithmically divergent at the $\l\ra 0$ end of the integration even in an abelian theory, and this is because the Coulomb potential in an abelian theory confines with a logarithmically rising potential.  The criterion that the infrared divergence in the self-energy is stronger than logarithmic is 
\beq
\lim_{\l \ra 0} \left\langle {\r(\l)  (\phi_\l|-\nabla^2|\phi_\l)  \over \l^{1-\e} }\right\rangle > 0 ~~\mbox{for some}~~ \e > 0
\label{condition}
\eeq
This condition involves the near-zero F-P eigenmodes, as well as the eigenvalues.  However, assuming that the eigenvalue spectrum is non-degenerate (apart from some rather special cases involving symmetric gauge-field configurations),
then at fixed ${g^2>0}$ each $\l_n$ is associated with a unique $(\bn,A)$, which in turn determines $\bp$.
Then $p^2 = p^2(\l)$ in the infinite volume limit and
\beq
\lim_{\l \ra 0}\left\langle {\r(\l)  (\phi_\l|-\nabla^2|\phi_\l)  \over \l^{1-\e}} \right\rangle \ge   
     \lim_{\l \ra 0}   \left\langle {\r(\l) p^2(\l) \over \l^{1-\e}} \right\rangle
\label{ie0}
\eeq
The proof of this inequality is given in the Appendix.  It follows that a sufficient condition for Coulomb confinement is
\beq
 \lim_{\l \ra 0}   \left\langle {\r(\l) p^2(\l) \over \l^{1-\e}} \right\rangle > 0 ~~\mbox{for some}~~ \e > 0
\label{criterion}
\eeq

\section{\label{FPE} The approach to the horizon}

   It was stated above that numerical simulations find local minima of $R$, which means, strictly speaking, that all of the eigenvalues of the F-P
operator are positive.  This statement has to be qualified a little.  Even apart from Gribov copies, the Coulomb and Landau gauge conditions
do not entirely fix the gauge, because if $U_\mu(x)$ satisfies the gauge condition, so does $GU_\mu(x) G^\dg$, where $G \in$ SU(N) is
any position-independent group element. This is a remnant global gauge symmetry, and it implies that at
any stationary point of $R$ there must be flat directions along the gauge orbit corresponding to zero modes 
of the F-P operator.  These are the trivial eigenmodes 
\beq
            \phi^a_{0,A}(x) = {1\over \sqrt{V}} \chi^a_A
\eeq
The statement that the F-P determinant is positive in the Gribov region really refers to the determinant of the operator in the subspace orthogonal to these trivial zero modes.

\begin{figure*}[t*]
\begin{center}
\subfigure[~Type I scenario.] 
{
    \label{conj1}
    \includegraphics[width=8truecm]{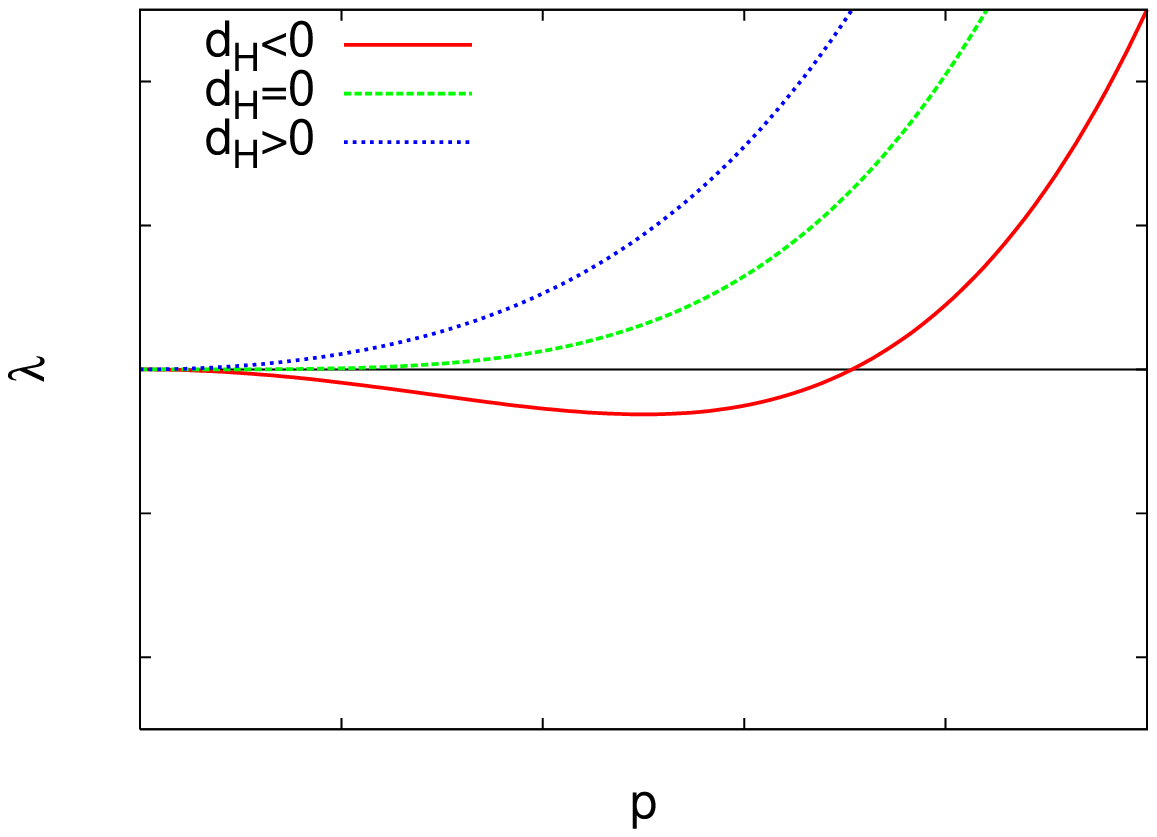}
}
\hspace{0.25cm}
\subfigure[~Type II scenario. ] 
{
    \label{conj2}
    \includegraphics[width=8truecm]{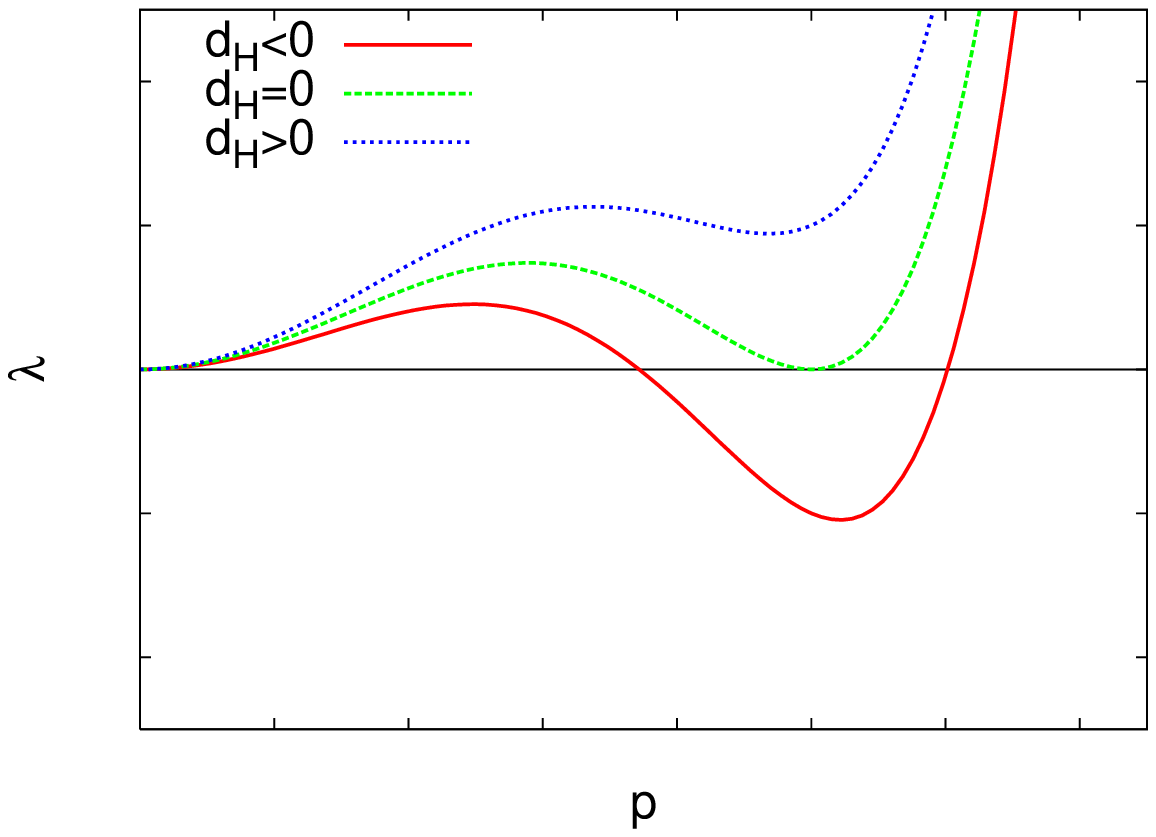}
}
\end{center}
\caption{Two scenarios for the behavior of the FP eigenvalue spectrum for gauge field configurations near the first Gribov horizon.
Just outside the Gribov region ($d_H<0$), there is a small interval of negative eigenvalues, which shrinks to a single eigenvalue at the horizon
($d_H=0$).  In the Type I scenario,  the interval of negative eigenvalues begins at $p=0$; at the horizon the nontrivial zero-mode is at $p=0$, and
at small $p$ the eigenvalues grow with a non-standard power $\l_p \sim p^{2+s}$.  For configurations inside the Gribov region ($d_H>0$) the growth
$\l_p \sim p^2$ is quadratic.  In the Type II scenario, for configurations just outside the Gribov region, the interval of negative eigenvalues does not include $p=0$, and at the Gribov horizon the non-trivial zero mode is at $|p|>0$.}  
\label{conj} 
\end{figure*}

    Outside the Gribov region, some of the non-trivial F-P eigenvalues become negative, which means that for configurations
which lie exactly on the Gribov horizon there must be at least one non-trivial zero eigenvalue.  However, in an infinite volume,
the converse is not necessarily true:  we cannot deduce, just from the fact that the spectrum of non-trivial eigenvalues begins
at zero, that the gauge field lies on the Gribov horizon.  Even in an abelian theory, which has no Gribov horizon, the spectrum of
the F-P operator $-\nabla^2$ in an infinite volume begins at $\l=0$. 

   Let us begin with $g^2=0$, i.e.\ a free-field theory, with the eigenvalues and eigenstates shown in \rf{zeroth}.  In this free case
we have\footnote{To compute $\rho(\l)$, begin with the volume measure in momentum space, proportional to $p^{d-1} dp$, and change
variables to $\l=\l(p)$ to arrive at $\rho(\l) d\l$.}
\beq
         \rho(\l) \propto \l^{(d-2)/2}  ~~~,~~~ (\phi_\l | -\nabla^2 |\phi_\l) = \l
\eeq
so that with an ultraviolet regulator, the Coulomb self-energy in $d+1$ dimensions is finite for all space dimensions $d \ge 3$, and
marginally divergent (divergent as $\log(L)$ as extension $L \ra \infty$) at $d=2$.  The latter divergence
is expected, since the Coulomb potential increases logarithmically in $2+1$ dimensions, so the question in 2+1 dimensions is
whether the condition in eq.\ \rf{condition} is satisfied for some $\e > 0$
    
   Outside the Gribov region, some of the non-trivial F-P eigenvalues become negative, and approaching the first Gribov horizon from 
the outside, the range of negative eigenvalues should shrink away. Right on the horizon there must exist
a non-trivial zero eigenvalue even for a finite spacetime volume. So let us imagine increasing 
$g^2$ away from zero, and also placing a constraint in 
the functional integral by introducing a dimensionful parameter $d_H$, and requiring that if $d_H>0$, the integration
is over gauge fields inside the Gribov region, lying a distance $d_H$ from the first Gribov horizon, while if $d_H<0$, the integration is
over gauge fields \emph{outside} the Gribov region, at a distance $|d_H|$ from the horizon.   Then
\bea
         \langle \l_p \rangle &=& \l^{(0)}_p + \langle \Delta \l_p \rangle 
\non \\
                              &=& p^2 (1 - F[g,p,d_H])
\eea
We use $p$ as an index because, in the infinite volume limit, it is better to replace the integer index $\bn$ by the continuous index $p$.  Also the expectation value of $\l_{\bp,A}$ can depend on neither the index $A$, since this would violate global color symmetry, nor on the direction of $\bp$, which would violate rotation invariance.
   
   If $g=0$ then $F=0$, but we may speculate on the behavior of $F[g,p,d_H]$ at $g^2>0$ as $d_H$ varies.  Suppose $F$ has the form, near $p=0$,
\bea
          F[g,p,d_H] &=& a[g,d_H] - b[g,d_H] p^s
\non \\
             & & \qquad + \text{higher powers of} ~ p
\eea
and $b[g,d_H]$ is positive for small $|d_H|$.  At $d_H > 0$ all non-trivial eigenvalues are positive, so it must be that $a[g,d_H]<1$ for small $p$.  Note that the eigenvalue spectrum in an infinite volume still starts at $\l=0$, even though the configurations are, by definition, off the Gribov horizon.  At $d_H<0$ some eigenvalues are negative, and if those are the eigenvalues near $p=0$, it means that $a[g,d_H]>1$.  The negative eigenvalues must just disappear at $d_H=0$, and this is obtained if $a[g,0]=1$ exactly.  
In this last case the subleading power of $p$ in $F[g,p,d_H]$ takes over,
and we have
\bea
           \l_p &\sim& p^{2+s}
\non \\
           \rho(\l) &\sim& \l^{(d-2-s)/(2+s)}
\label{horizon}
\eea
This a qualitative change in the low-lying F-P spectrum, compared to the behavior inside the Gribov region, and the sufficient condition \rf{criterion} for Coulomb confinement is satisfied if
\beq
        2s + 2 > d
\label{cc-s}
\eeq
Inside the Gribov region, at $p\ra 0$, the spectrum
is simply a rescaling of the zeroth-order spectrum
\beq
         \l_p = (1-a[g,d_H]) p^2
\eeq
and, in the case of Coulomb gauge, the Coulomb self-energy is finite.
The conjectured behavior of $\langle \l_p \rangle$ vs.\ $p$, for $d_H$ positive, negative, and zero, is sketched in Fig.\ \ref{conj1}. 

    But the scenario just outlined is not the only possible behavior near the horizon.  Consider, in particular, the case that $b[g,d_H]$ is negative for small $|d_H|$.  Then we have
\beq
        \langle \l_p \rangle = (1 - a[g,d_H]) p^2 -  \Bigl| b[g,d_H]\Bigr| p^{2+s} + \text{higher powers of} ~ p
\eeq
and it is possible that $\langle \l_p \rangle$ is positive near $p=0$ where the $p^2$ term dominates, but negative in some finite region away from $p=0$.  The conjectured behavior in this case, for positive, negative, and vanishing $d_H$, is indicated in Fig.\ \ref{conj2}, and in this case we would still have $\l_p \sim p^2$ at the horizon, for small $p^2$.

         Of course, quantization in Coulomb and Landau gauge does not involve setting $d_H$ to some definite value.  What \emph{is} required, however, is a constraint on the range of functional integration to lie within the first Gribov horizon.  If it is true that entropy dominates due to the
high dimensionality of the configuration space, and almost all of the volume of the Gribov region is concentrated at the horizon, then only lattice configurations at or very near the horizon will contribute to vacuum expectation values in Coulomb and Landau gauge,  just as if the constraint $d_H=0$ were imposed.

\section{Perturbative evaluation of the F-P spectrum}     
            
     The possible spectra shown in Figs.\ \ref{conj1} and \ref{conj2} are pure speculation at this point, but it is interesting, and somewhat in the spirit of Gribov's original work \cite{Gribov}, to see how far we can go in understanding the F-P spectrum with ordinary perturbation theory.

   Let us begin with lattice SU(2) gauge theory in either $d+1$ spacetime dimensions (Coulomb gauge) or $d$ spacetime dimensions
(Landau gauge), starting on a finite $d$-dimensional volume $V$ and taking the infinite volume $V\ra \infty$ and
lattice spacing $a\ra 0$ limits at the end.   The F-P operator on the lattice is given by \cite{GOZ2}
\bea
           M_{xy}^{ab} &=&  (K_0)^{ab}_{xy} + (K_1)^{ab}_{xy} + (M_1)^{ab}_{xy}
\non \\
           (K_0)^{ab}_{xy} &=& \d^{ab} \sum_i (2\d_{xy} - \d_{x+\hi,y} - \d_{x-\hi,y})
\non \\
           (K_1)^{ab}_{xy} &=& \oh g \e^{acb} \sum_i \left[ -A_i^c(x) \d_{x+\hi,y} + A^c_i(y) \d_{x-\hi,y} \right]
\non \\
           (M_1)^{ab}_{xy} &=& - \d^{ab} \sum_i \Bigl\{ \d_{xy} \left[(1- \oh \mbox{Tr}U_i(x)) + (1- \oh \mbox{Tr}U_i(x-\hi))\right] 
\non \\
             & & \left. - \d_{x+\hi,y}(1- \oh \mbox{Tr}U_i(x))  - \d_{x-\hi,y}(1- \oh \mbox{Tr}U_i(y)) \right\}  
\non \\
\eea
where
\beq 
                A_j^a =  {1\over 2ig} \mbox{Tr}[\s_a(U_j(x) - U^\dg_j(x))]
\eeq
The dimensionless lattice coupling $g_L$ is related to the gauge coupling $g$ by $g^2_L = a^{4-D}g^2$, where $a$ is the lattice spacing and $D$ is the spacetime dimension.
The eigenvalues and eigenvectors of $K_0$ are those shown in eq.\ \rf{zeroth}. The operator $M_1$ vanishes in the 
continuum limit, so I will just ignore it in what follows,  and treat $K_1$ as the only
perturbation to $K_0$.   Lattice Fourier transforms will be defined symmetrically
\bea
             A_i^a(x) &=& {1\over \sqrt{V}} \sum_k \tA_i^a(k) e^{ikx}
\non \\
             \tA_i^a(k) &=& {1\over \sqrt{V}} \sum_x A_i^a(x) e^{-ikx}
\label{FT}
\eea

     The first-order correction to $\l^{(0)}_p$ is 
\bea
              \D \l^{(1)}_{p,A} &=& \langle p,A| K_1 |p,A\rangle
\non \\
            &=& {1\over V} \sum_x \sum_y e^{-ipx} \chi_A^a (K_1)^{ac}_{xy} e^{ipy} \chi_A^c
\non \\
            &=& \oh g \chi_A^a \e^{abc} \chi_A^b \sum_i \sum_x  {1\over V} \left[-A_i^b(x)e^{ip_i} + A_i^b(x-\hat{i})e^{-ip_i} \right]
\non \\
            &=& -ig \chi_A^a \e^{abc} \chi_A^b {1\over \sqrt{V}} \sum_i \tA_i^b(0) \sin(p_i)
\label{first}
\eea
Now, according to the above definition of the lattice Fourier transform, the lattice $A$-field at zero momentum is
\beq
               \tA_i^b(0) = {1\over \sqrt{V}} \sum_x A_i^b(x)
\eeq
with $-2/g < A^b_i(x) < 2/g$.  Then suppose  that the lattice $A$-field in Coulomb or Landau
gauge has a finite correlation length $l$.  This implies
\beq
           \sum_x A_i^b(x) \sim \pm \sqrt{V \over l^d} l^d \A
\eeq
where $\A$ is the average value of $A_i^b$ in a hypercubic region of volume $l^d$. Then, because of the factor of $1/\sqrt{V}$ in \rf{first}, the first-order correction to $\l_p^0$ vanishes in the infinite volume
limit.  Of course, the first-order contribution vanishes even in a finite volume upon taking the expectation value, since 
$\langle \tA_i^b(0)\rangle = 0$.
 
     At second order
\beq
          \D \l_{p,A} = \sum_{k,B} { |( k,B | K_1 | p,A) |^2 \over \l^0_p - \l^0_k }
\eeq
where
\bea
 \lefteqn{( k,B | K_1 | p,A)}
 \non \\
    &=&  \oh g \chi_B^a \e^{abc} \chi_A^c \sum_i {1\over V} \sum_x e^{-ikx} 
             \left( -A_i^b(x) e^{ip(x+\hi)} \right.
\non \\ & & \left. \qquad + A_i^b(x-\hi) e^{ip(x-\hi)} \right)
\non \\
   &=& \oh g \chi_B^a \e^{abc} \chi_A^c {1\over \sqrt{V}}  \sum_i A_i^b(k-p) \left( -e^{ip_i} + e^{-ik_i} \right)
\non \\
\eea
Then
\bea
\D \l_{p,A} &=& \oq g^2 \sum_B (\chi_A^a \e^{abc} \chi_B^c) (\chi_A^d \e^{def}\chi_B^f) 
                           {1\over V} \sum_k {1\over \l^0_p - \l^0_k}
\non \\
   & & \qquad \times \sum_{ij} \tA_i^b(k-p) \tA_j^e(p-k) 
\non \\
   & &\qquad \times \left( -e^{ip_i} + e^{-ik_i} \right)\left( -e^{-ip_j} + e^{ik_j} \right)
\label{DL1}
\eea

     In preparation for taking the continuum limit, we need to indicate powers of the lattice spacing explicitly.
Let
\beq
      \D \l = a^2 \D \l' ~~,~~ p = a p'  ~~,~~  A_i^c(x) = a A'^c_i(x)
\eeq
where the primed quantities have the standard engineering dimensions of these quantities in the continuum formulation.
We also have, using $\D k' = 2\pi /(La)$,
\bea
    {1\over V} \sum_k &=& {1\over L^d} {1\over (\D k')^d} \sum_k (\D k')^d
\non \\
               &=& a^d \sum_k \left( {\D k' \over 2\pi}\right)^d
\eea
Inserting these identities into \rf{DL1}
\bea
 \D \l'_{p,A} &=& {1\over a^2} {g^2 \over 4} \sum_B (\chi_A^a \e^{abc} \chi_B^c) (\chi_A^d \e^{def}\chi_B^f) 
\non \\
            & & \times  a^d \sum_k \left( {\D k' \over 2\pi}\right)^d {1 \over a^2(\l'^{(0)}_p - \l'^{(0)}_k)}
\non \\
    & &  \sum_{ij} \tA^b_i(k-p) \tA^e_j(p-k) 
\non \\
    & &\times \left( -e^{ip'_ia} + e^{-ik'_ia} \right)\left( -e^{-ip'_ja} + e^{ik'_ja} \right)
\eea

     Now we take the vacuum expectation value of $\D \l'$, noting that 
\beq
          \langle \tA^b_i(k) \tA^c(-k) \rangle = a^{2-d} \d^{bc} D_{ij}(k')
\eeq
where, in Landau gauge, $D_{ij}(k')$ is the full (i.e.\ dressed) gluon propagator.  In Coulomb gauge it is the spatial Fourier transform of the
full, equal-times gluon propagator. 
This gives 
\bea
 \langle \D \l'_p \rangle &=& {1\over a^2} {g^2 \over 4} \sum_B (\chi_A^a \e^{abc} \chi_B^c) (\chi_A^d \e^{dbf}\chi_B^f) 
\non \\
            & & \times \sum_k \left( {\D k' \over 2\pi}\right)^d {1 \over \l'^{(0)}_p - \l'^{(0)}_k} \sum_{ij}   D_{ij}(p'-k')
\non \\
    & & \times  \left( -e^{ip'_ia} + e^{-ik'_ia} \right)\left( -e^{-ip'_ja} + e^{ik'_ja} \right)
\eea
At this point we can take the continuum limit, and make use of the transversality property $q_i D_{ij}(q)=0$ of the gluon propagator, to obtain\footnote{We have ignored, in lattice regularization, the case that $\l'^{(0)}_p = \l'^{(0)}_k$, which would have to be handled by degenerate perturbation theory.  This case is zero measure in the continuum limit, and will not require special treatment.}
\bea
\langle \D \l'_p \rangle &=&  g^2  \sum_B (\chi_A^a \e^{abc} \chi_B^c) (\chi_A^d \e^{dbf}\chi_B^f) 
\non \\
            & & \times \int {d^d k' \over (2\pi)^d} {1 \over p'^2 - k'^2}  p'_i p'_j D_{ij}(p'-k') 
\eea
 
     The primes, having served their purpose, will now be dropped.  It is understood that the unprimed quantities
now have their standard engineering dimensions.
     
     Using the competeness property
\beq
          \sum_B \chi_B^c \chi_B^f = \d^{cf}
\eeq
we sum over the color indices, which just gives an overall factor of two.  The result is
 \beq
     \langle \D \l_p \rangle = -2 g^2 p_i p_j \int {d^d k \over (2\pi)^d} {1\over k^2 - p^2} D_{ij}(p-k)
\eeq
(note the interchange of $k^2$ and $p^2$ in the denominator).
Changing variables to $q=p-k$, and writing
\beq
        D_{ij}(q) = \left(\d_{ij} - {q_i q_j \over q^2} \right)  D(q)
\label{general_prop}
\eeq
gives
\bea
    \langle \D \l_p \rangle &=& -2 g^2 \int {d^d q \over (2\pi)^d} ~ {D(q) \over (p-q)^2 - p^2} 
\non \\
       & & \quad \times \left(p^2 - {(p\cdot q)^2 \over q^2} \right)
\label{generalD}
\eea

   We now go to $d$-dimensional spherical coordinates
\beq
       \int d^dq = A_{d-1} \int_0^\infty dq ~ q^{d-1} \int_0^{\pi} \sin^{d-2}\th
\eeq
where
\beq
         A_{d-1} = {2\pi^{(d-1)/2} \over \Gamma\left({d-1\over 2}\right) }
\eeq
Define
\beq
\tD(q) = q^{d-2} D(q) 
\eeq
and
\beq
R_d = {2A_{d-1} \over (2\pi)^d} = \left\{ \begin{array}{cc}
                                           1/\pi^2 & d=2 \cr
                                           1/(2\pi^2) & d=3 \cr
                                           1/(6\pi^3) & d=4 
                                          \end{array} \right.
\eeq
Then
\begin{widetext}
\bea
  \langle \D \l_p \rangle &=& -g^2 R_d \int_0^\infty dq q^{d-1 } \int_0^\pi d\th ~ \sin^{d-2}\th 
  (1-\cos^2 \th) {1\over q^2-2pq\cos\th}  D(q)  p^2
\non \\
      &=& -g^2 R_d \int_0^\pi d\th ~ \sin^{d-2}\th (1-\cos^2 \th)
    \int_0^\infty dq {1\over q-2p\cos\th} q^{d-2} D(q)  p^2
\non \\
      &=& -g^2 R_d  p^2 (I_1 + I_2)
\eea
where
\bea
I_1 &=& \int_0^{\pi/2} d\th ~ \sin^{d-2}\th (1-\cos^2\th) \int_0^\infty dq {1\over q-2p\cos\th} \tD(q)
\non \\
I_2 &=& \int_{\pi/2}^\pi d\th ~ \sin^{d-2}\th (1-\cos^2\th) \int_0^\infty dq {1\over q-2p\cos\th} \tD(q)
\non \\
\eea
Make the change of variables $\th \ra \pi - \th$ in $I_2$
\beq
I_2 = \int_0^{\pi/2} d\th ~ \sin^{d-2}\th (1-\cos^2\th) \int_0^\infty dq {1\over q+2p\cos\th} \tD(q)
\eeq
Then, in $I_1$, it is useful to rewrite the integral over momenta $q$
\bea
\int_0^\infty dq {\tD(q) \over q - 2p\cos\th} &=& \left\{ \int_0^{2p\cos\th} + \int_{2p\cos\th}^{4p\cos\th}
   + \int_{4p\cos\th}^\infty \right\} dq {\tD(q) \over q - 2p\cos\th}
\non \\
&=& \int_0^{2p\cos\th} dq {\tD(q) \over q - 2p\cos\th} + \int_0^{2p\cos\th} dq {\tD(4p\cos\th - q) \over 2p\cos\th - q}
       + \int_0^\infty dq {\tD(4p\cos\th + q) \over 2p\cos\th + q}
\non \\
&=& \int_0^{2p\cos\th} dq {1 \over 2p\cos\th - q}[\tD(4p\cos\th - q) - \tD(q)]
        + \int_0^\infty dq {\tD(4p\cos\th + q) \over 2p\cos\th + q}
\eea
Altogether, we have to second order
\beq
    \langle \l_p \rangle = \l_p^{(0)} + \langle \D \l_p \rangle
                 = p^2 \Bigl(1 - g^2 R_d  I[p,m,\a] \Bigr)
\label{pt2}
\eeq
where
\bea
 I[p,m,\a]  &=&  \int_0^{\pi/2} d\th ~ \sin^{d-2}\th (1-\cos^2\th) \left\{ \int_0^\infty dq 
{1 \over q +2p\cos\th}[\tD(4p\cos\th + q) + \tD(q)] \right.
\non \\
   & & \qquad \left. + \int_0^{2p\cos\th} dq {1 \over 2p\cos\th - q} [\tD(4p\cos\th - q) - \tD(q)] \right\}
\label{I}
\eea
\end{widetext}
The $m,\a$ in $I[p,m,\a]$ are constants I will use to parametrize the transverse gluon propagator $D(q)$.

\section{An ansatz for the gluon Propagator}

   The gluon propagators $D_{ij}$ in Coulomb and Landau gauges are transverse with respect to spatial momenta ${\mathbf q}$ in $d+1$ dimensions,
and spacetime momenta $q^\m$ in $d$ Euclidean dimensions, respectively.  Therefore these propagators have the form shown in \rf{general_prop}.   In a free theory
\beq
        D(q) = \left\{ \begin{array}{cl}
                       1/(2q) & \text{Coulomb gauge} \cr \cr
                        1/q^2 & \text{Landau gauge}
                       \end{array} \right.
\label{Dfree}
\eeq
where the propagator in Coulomb gauge is at equal times, with $q$ the space (rather than spacetime) momentum.
The behavior \rf{Dfree} is expected at high momenta, but it is certainly not correct at low momenta, as seen from lattice Monte Carlo simulations.  In Landau gauge, the current evidence is that $D(0)$ is finite and non-zero at $q=0$ in three and four dimensions \cite{Brazil,Berlin}, while  $D(q) \ra 0$ in two dimensions \cite{Maas}.  In Coulomb gauge it appears that $D(q) \ra 0$ in four dimensions \cite{coulomb}.
  
   In order to allow for non-singular power behavior in the transverse gluon propagator as $p\ra 0$, I will adopt the ansatz
that  
\beq
        D(q) = {1 \over 2 \sqrt{q^2 + m^{2+\a}/q^\a}} 
\label{D-coul}
\eeq
in Coulomb gauge, and
\beq
        D(q) = {1 \over q^2 + m^{2+\a}/q^\a} 
\label{D-lan}
\eeq
in Landau gauge.  Gribov's proposal for the gluon propagator in these cases corresponds to $\a=2$.  The propagators go over to free-field behavior as $q\ra \infty$.

\begin{figure}[h!]
\centerline{\scalebox{0.7}{\includegraphics{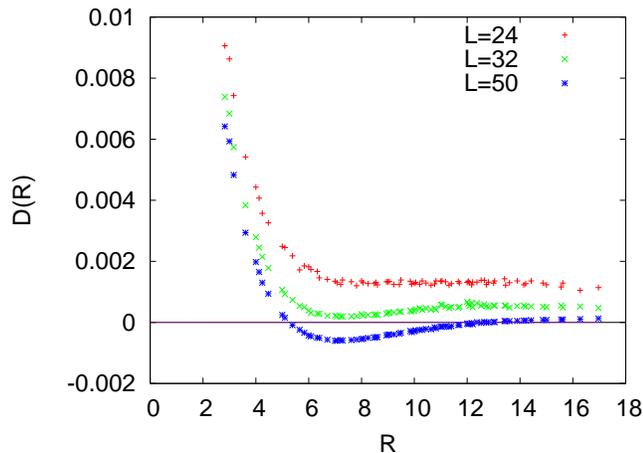}}}
\caption{Equal-times Coulomb-gauge gluon propagator in 2+1 dimensions, at $\b=6$ and
$L^3$ lattice volume, for $L=24,32,50$.}
\label{prop}
\end{figure}

    I am not aware of any lattice Monte Carlo computation of the transverse gluon propagator in Coulomb gauge in
$d=3$ dimensions, in position space.  In Fig.\ \ref{prop} I show data for $D(R)$ obtained from
the equal times correlator
\beq
      \langle \mbox{Tr}[\texttt{A}_j(\bx,t) \texttt{A}_j(\by,t)]
\eeq
of gluon fields 
\beq
       \texttt{A}_j(\bx,t) =  {1\over 2i}(U_j(\bx,t) - U^\dg_j(\bx,t))
\eeq
on the lattice.  The correlator is calculated via lattice Monte Carlo with an SU(2) Wilson action, 
on an $L^3$ lattice volume at coupling $\beta=6$ and $L=24,32,50$, with the equal-times correlator computed after transforming
the gauge fields to Coulomb gauge.
Note that as the lattice volume increases, the gluon propagator develops a ``dip'' and actually 
becomes negative at the larger $R$ values. This behavior appears to rule out
$\a=0$, in which the propagator should be everywhere positive.  A reliable computation of $D(q)$ as $q \ra 0$ will 
probably require a large-scale lattice calculation, as has been done for the Landau gauge.

\begin{figure*}[t!]
\centerline{\scalebox{0.5}{\includegraphics{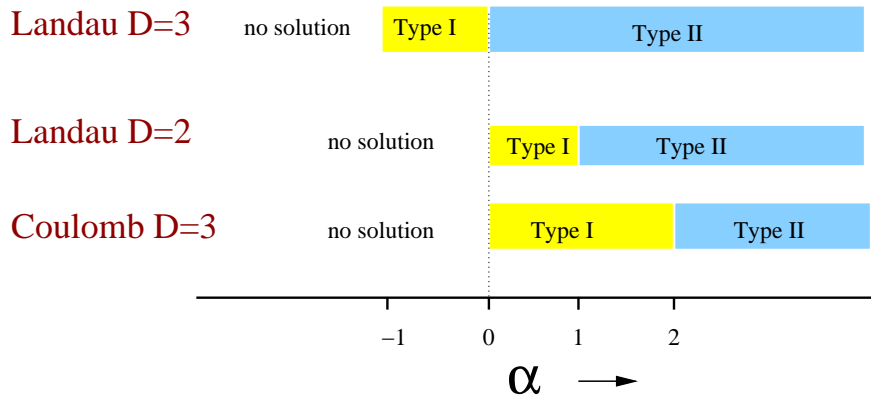}}}
\caption{Summary  of the qualitative behavior of the low-lying F-P spectra, according to 2nd order perturbation theory,
for Landau and Coulomb gauges in D=2 and 3 dimensions.   The sketch illustrates how the behavior of the F-P spectra
depends on the assumed infrared behavior of the gluon propagator, which is parametrized by the exponent $\a$.}
\label{fp}
\end{figure*}

\section{Results for F-P Spectra}

       In section \ref{FPE} I introduced a parameter $d_H$ to control the approach to the first Gribov horizon, and speculated on the low-$p$ behavior of $\l_p$ as the horizon is approached.  In the perturbative calculation, the mass parameter $m$ in the gluon propagator plays essentially the same role as $d_H$.   Note that in dimensions lower than 3+1, where $I[p,m,\a]$ is convergent, the coupling
$g^2$ is dimensionful, and we may as well choose units such that $g^2=1$.   Then
\beq
        \langle \l_p \rangle = p^2 \Bigl((1 - R_d I[p,m,\a] \Bigr)
\label{pt2a}
\eeq
Expanding $I[p,m,\a]$ in leading powers of $p$ near $p=0$, we have
\bea
       R_d I[p,m,\a] &=& a[m,\a] - b[m,\a] p^s 
\non \\
             & & \qquad + \mbox{higher powers of $p$}
\eea
in which case
\bea
      \langle \l_p \rangle &=& (1-a[m,\a]) p^2 + b[m,\a] p^{2+s} 
\non \\
             & & \qquad +  \mbox{higher powers of $p$} 
\label{general}
\eea
Suppose, for a given $\a$, it is possible to find a critical value $m=m_c$ such that $a[m_c,\a]=1$ and $b[m,\a]>0$.
In that case we have the Type I scenario conjectured in Fig.\  \ref{conj1} above; i.e.
\begin{enumerate}
\item $m<m_c$ and $a[m,\a] > 1$:   The low-lying F-P eigenvalue spectrum has a range of negative eigenvalues, starting at $p=0$. We interpret this to mean that the transverse gluon propagator, which determines the spectrum at second order, is determined by configurations outside the Gribov region.
\item $m=m_c$ and $a[m_c,\a]=1$: The region of negative eigenvalues just disappears, and $\l_p \sim p^{2+s}$.  This is the case of particular
interest, where the gluon propagator is derived from configurations which mainly lie right on the Gribov horizon. 
\item $m>m_c$ and $a[m_c,\a] < 1$.  In this case the low-lying spectrum $\l_p = (1-a[m,\a])p^2$ is just a rescaling of the free-field spectrum, 
and the gluon propagator is derived from configurations inside the Gribov region.
\end{enumerate}

\begin{figure*}[tbh]
\begin{center}
\subfigure[] 
{
    \label{lowmass}
    \includegraphics[width=8truecm]{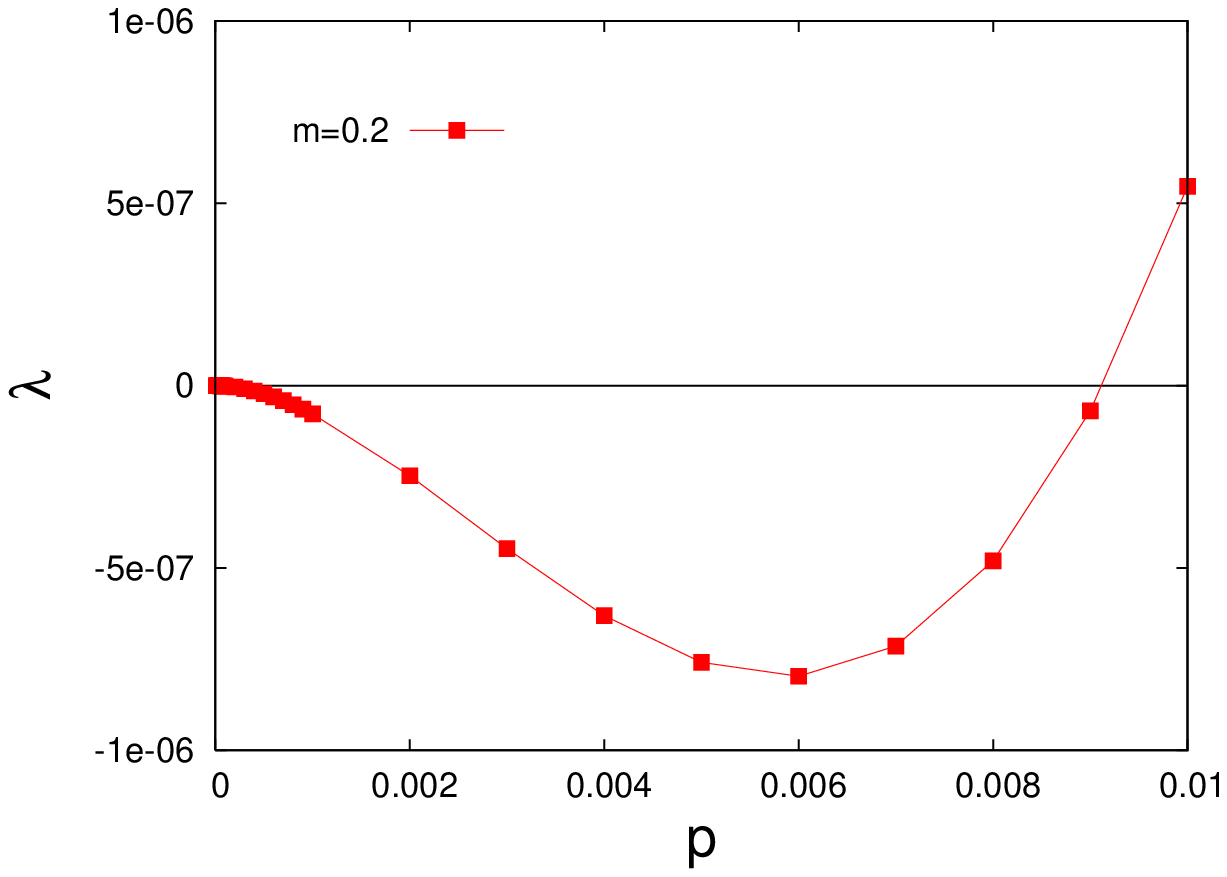}
}
\hspace{0.25cm}
\subfigure[] 
{
    \label{mcrit}
    \includegraphics[width=8truecm]{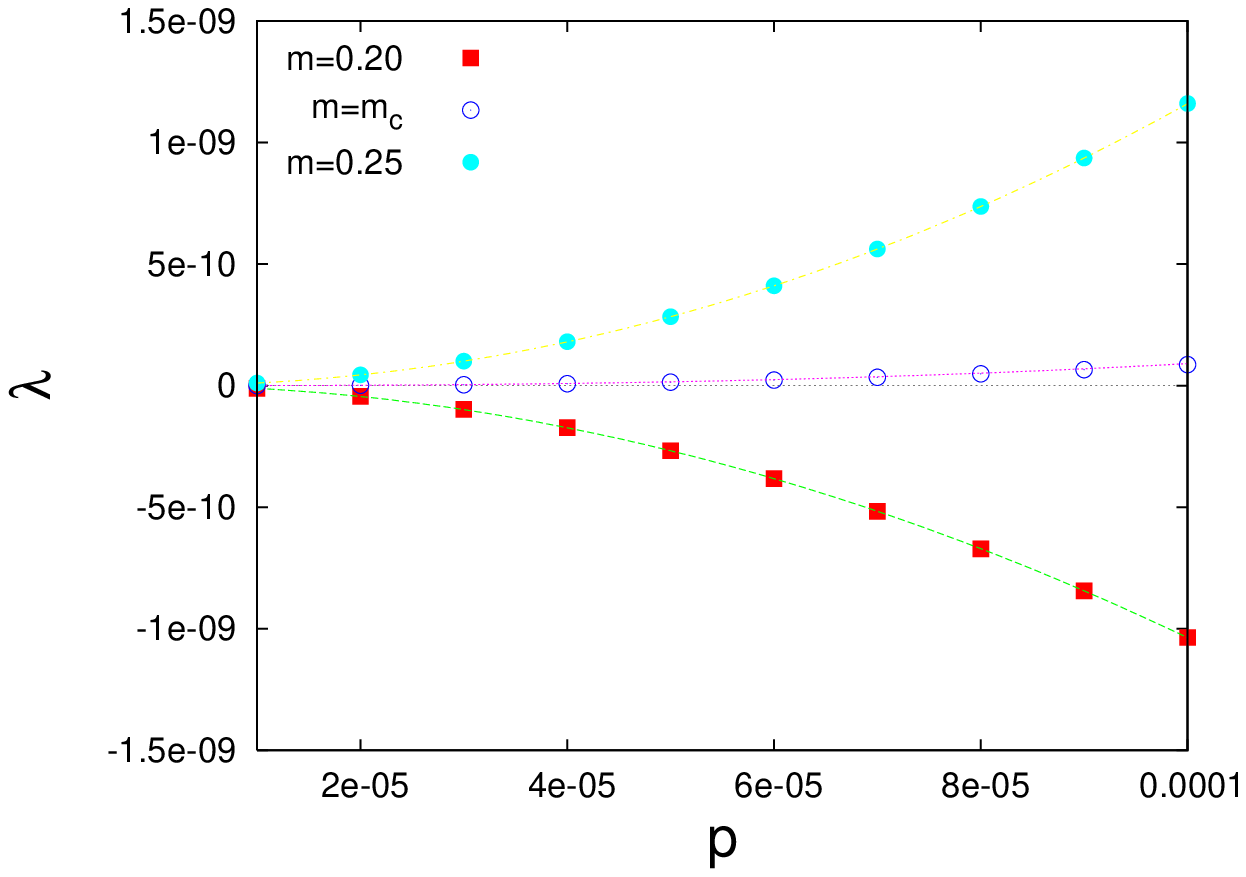}
}
\end{center}
\caption{F-P spectra at $\a=1$. (a) $m=0.20<m_c$.  There is an interval of negative eigenvalues in the region $0<p<0.009$. 
(b) $\l_p$ at low $p$, for $m=0.20 < m_c$, $m=m_c=0.2228$, and $m=0.25>m_c$.}  
\label{masses} 
\end{figure*}

\begin{figure}[h!]
\centerline{\includegraphics[width=8truecm]{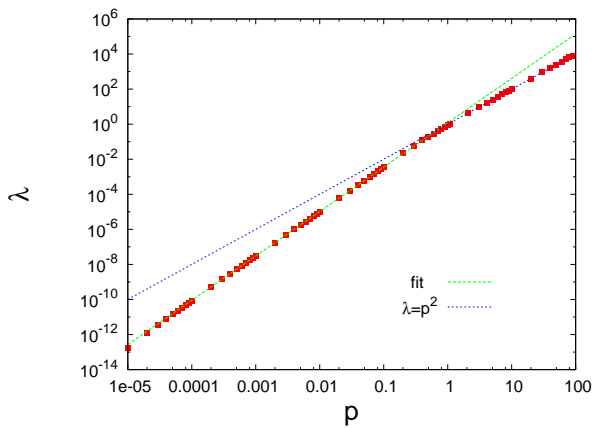}}
\caption{Log-log plot of the spectrum of the Fadeev-Popov operator, for
$\a=1$ at the critical $m_c=0.223$. A best fit at $p<1$ yields $\l_p = 1.21 p^{2.53}$.}
\label{alf10}
\end{figure}

\begin{figure}[h!]
\centerline{\includegraphics[width=8truecm]{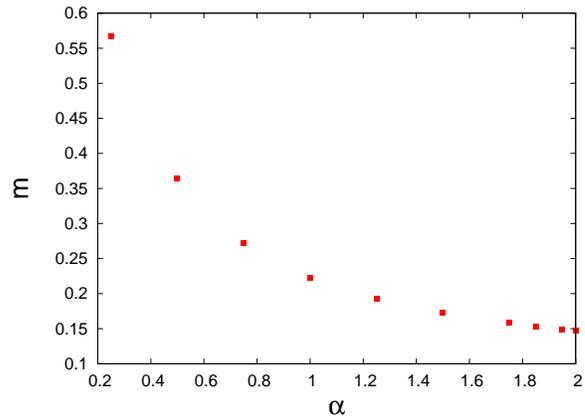}}
\caption{Critical value $m_c$ for the mass parameter in the transverse
gluon propagator, vs.\ the power $\a$.}
\label{mass}
\end{figure}

 It should be noted at this point that the Type I scenario is in some ways reminiscent of the Dyson-Schwinger approach, and indeed eq.\ \rf{pt2} 
resembles the Dyson-Schwinger equation for the ghost propagator in covariant gauges (see, e.g.,
Fischer \cite{Christian}).  Of course these equations are not the same; \rf{pt2a} is an equation for the expectation value of F-P eigenvalues, not the inverse ghost propagator, and it is derived from a perturbative expansion, not the Dyson-Schwinger equations.  Nevertheless, the scaling  solution   \cite{scaling} is obtained from the Dyson-Schwinger equation by tuning a coupling so that the bare inverse ghost propagator in that
equation is exactly cancelled by another term.  In the absence of this tuning, the decoupling solution \cite{decoupling} is obtained.   Similarly, in our approach, a mass parameter is tuned to exactly cancel the $p^2$ term in the eigenvalue spectrum, resulting  
in an enhanced density of near-zero eigenmodes.   The motivation for the tuning in our case is to study the F-P spectrum at
the Gribov horizon, which is only relevant to physics if, in fact, the functional integral over the Gribov region is dominated by horizon configurations.

    The Type II scenario is obtained if $b[m,\a]$ is negative when $a[m,\a]=1$, in which case there is still a range of negative eigenvalues, so 
this value of $m$ is not the critical value.  The critical value, corresponding to the horizon, is obtained at a value $m=m_c$ where $a[m,\a]<1$, such that the function
\beq
           \langle \l_p \rangle \approx \Bigl(1-a[m_c,\a]\Bigr) p^2 - \Bigl|b[m_c,\a]\Bigr| p^{r} + c[m_c,\a] p^q 
\eeq
approximating $\langle \l_p \rangle$ at small $p$ has a zero value, but no negative values, for one choice of $p\ne 0$.   In this case the horizon does not alter the power dependence $\l_p \sim p^2$ near $p=0$.

    Both the Type I and Type II scenarios assume that
\beq
             a[m,\a] = R_d I[p=0,m,\a]
\eeq
is finite.  This is not necessarily the case, however, and it is easy to check that $I[0,m,\a]$ is divergent at all 
$\a \le 0$ for Landau gauge in two dimensions and Coulomb gauge in three dimensions, and is divergent for all $\a \le -1$
for Landau gauge in three dimensions.  There is no choice of $m$, for those choices of $\a$, which completely eliminates negative F-P eigenvalues.
This will be referred to as the ``no solution" case.

\begin{figure*}[tbh]
\begin{center}
\subfigure[] 
{
    \label{power}
    \includegraphics[width=8truecm]{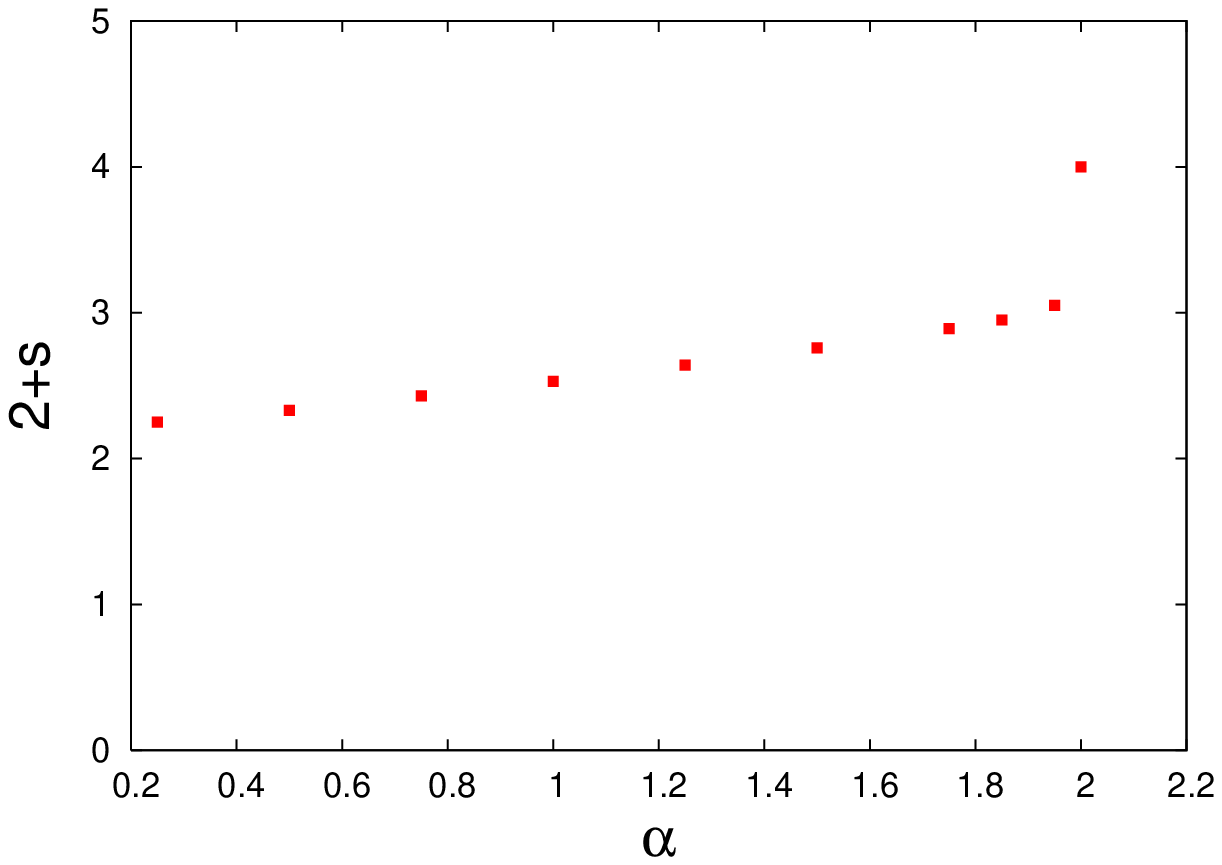}
}
\hspace{0.25cm}
\subfigure[] 
{
    \label{coeff}
    \includegraphics[width=8truecm]{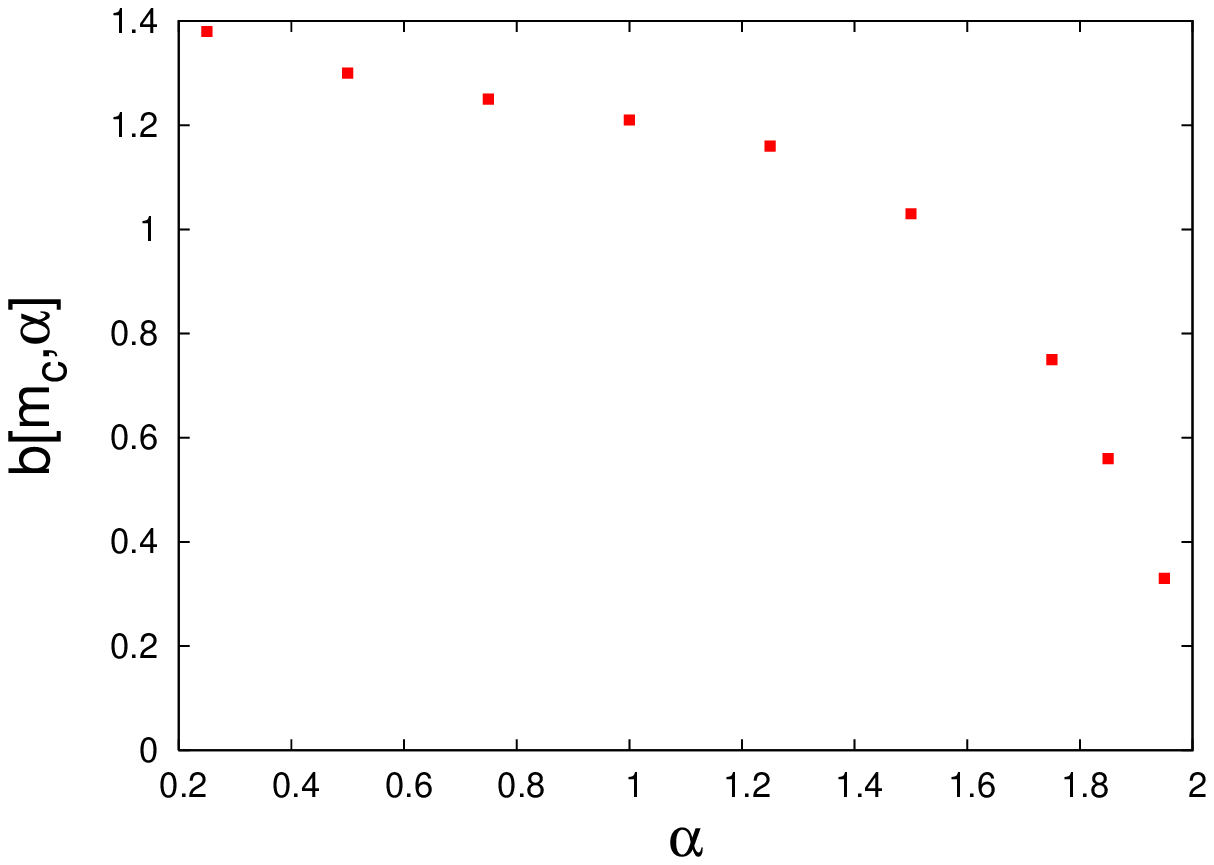}
}
\end{center}
\caption{(a) Exponent $2+s$ vs.\ $\a$; and (b) coefficient $b$ vs $\a$; for the power-law behavior $\l_p=b p^{2+s}$ at the critical
mass parameter $m=m_c(\a)$, for the Coulomb gauge F-P spectrum in 2+1 dimensions.  The sudden rise to $2+s=4$ at $\a=2$ is correlated with $b\ra 0$}  
\label{fit} 
\end{figure*}

      In order to determine which scenario is realized, at each choice of $\a$ for which $I[0,m,\a]$ is finite, it is necessary to calculate 
$I[p,m,\a]$ numerically.  The result, for Coulomb gauge in three dimensions, and Landau gauge in two and three dimensions, is indicated schematically in Fig.\ \ref{fp}.  To illustrate how these results are obtained, we consider in particular the case of $\a=1$ for Coulomb and Landau gauges in
three dimensions (i.e.\ $d=2$ for Coulomb, and $d=3$ for Landau).   We begin with Coulomb gauge (Figs.\ \ref{masses}-\ref{fit}).  Figure \ref{lowmass} shows the low-lying F-P spectrum at $\a=1$ and $m=0.20<m_c$, and it is clear that there is a region of negative eigenvalues starting at $p=0$. As $m$ is increased, the region of negative eigenvalues shrinks in size, until at a critical value $m=m_c(\a)$ the interval of negative eigenvalues just vanishes.   Figure \ref{mcrit} displays the low-lying spectrum just below, at, and just above the critical mass at $\a=1$, which is $m_c=0.2228$.  
At $m \ne m_c$, $\l_p$ is proportional to $p^2$ near $p=0$, with a proportionality constant which is positive or negative, depending on
whether $m$ is greater or less than $m_c$.  But precisely at $m=m_c$, we find that $\l_p \propto p^{2+s}$, with $s=s(\a) > 0$.
Fig.\ \ref{alf10} is a log-log plot of $\l_p$ vs.\ $p$ over a large range of $p$, at $\a=1$ and $m_c=0.223$.  For the range $0<p<1$, we can
determine that $s=0.53$ in this case, and $\l_p \approx 1.21 p^{2.53}$ at small $p$.  At around  $p\equiv |\bp |=1$ (in units $g^2=1$), the power behavior shifts to the free case, $\l_p=p^2$, and continues that way for all higher $p$, as expected.   This is an example of the Type I scenario.

   The next question is how $m_c$ and $2+s$ change as $\a$ is varied.  As already noted, we must choose $\a > 0$ to reach the horizon, which means that $D(0)=0$, and therefore the transverse gluon propagator must vanish at zero momentum for Coulomb gauge in 2+1 dimensions, and for Landau gauge in 1+1 dimensions.  As $\a \ra 0^+$, the increasingly singular behavior of the integrand in $I[p,m,\a]$ must be countered by an increasingly large value of $m_c$, in order to satisy $a[m_c],\a]=1$.  A plot of $m_c$ vs.\ $\a$ is shown in Fig.\ \ref{mass}.  

    The power behavior $\l_p = b p^{2+s}$ in the low-lying spectrum is crucial for Coulomb confinement, and the 
exponent $2+s$ vs.\ $\a$, obtained at $m=m_c$ is shown in Fig.\ \ref{power}.  In 2+1 dimensions the condition for Coulomb confinement (beyond the
marginal divergence of the free theory) is that $s>0$, which is seen to hold throughout the range shown.   

    We also see that there is a sudden jump in $s$ from roughly $s=1$ to $s=2$ at $\a=2$.  This is where the transition from Type I to Type II behavior takes place.  As $\a\ra 2$ the coefficient $b[m_c,\a]$
approaches zero (cf.\ Fig. \ref{coeff}) and then changes sign.  Exactly at $\a=2$, where $b[m_c,\a]=0$, the term which has the next 
higher power in $p$ takes over, accounting for the sudden jump in $s$.       

\begin{figure}[h!]
\centerline{\scalebox{0.7}{\includegraphics{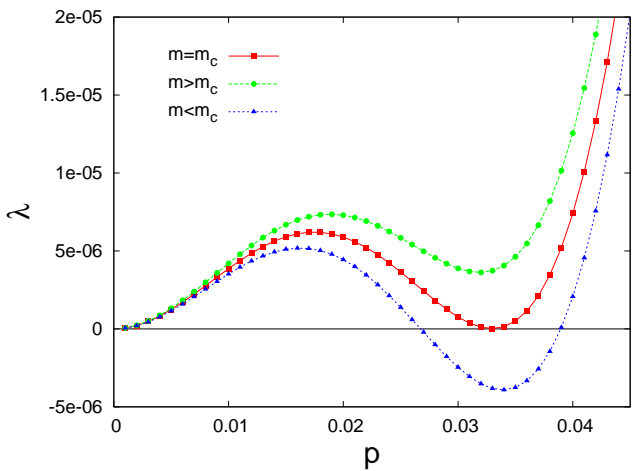}}}
\caption{The low-lying F-P eigenvalue spectra near the Gribov horizon, for Landau gauge in D=3 spacetime dimensions and $\a=1$, according to
second-order perturbation theory.  This is an example of the Type II scenario.}
\label{mcritL}
\end{figure}

     Landau gauge in three dimensions, at $\a=1$, furnishes an example of the Type II scenario.  The F-P spectrum at small $p$ is shown in Fig.\
\ref{mcritL} for the mass parameter above ($m=0.087$), below ($m=0.086$), and equal $m=m_c=0.08644$ to the critical value.

\section{Conclusions}

     If the integration over gauge fields is dominated by configurations on or near the first Gribov horizon, then the lowest non-trivial F-P eigenvalue
must be very close or equal to zero, even in a finite spacetime volume.  The main finding of the perturbative treatment presented here is that
if there is, in fact, a non-trivial zero mode,  and the F-P eigenvalues are labeled by the lattice momenta, then this non-trivial zero mode may occur at either zero momentum (Type I scenario) or non-zero momenta (Type II scenario), depending on the infrared behavior of the gluon propagator.
While the spectrum of F-P eigenvalues does not translate directly into a prediction for the behavior of the ghost propagator (because the
momentum behavior of the F-P eigenmodes must also be taken into account),
it is natural to conjecture that the Type I scenario is associated with an infrared singular ghost dressing function, as in Coulomb gauge, while the
Type II scenario corresponds to a finite ghost dressing function, as appears to be the case in Landau gauge.  This would most likely be the
case if $|\phi_{pA}(k)|^2$ is narrowly peaked around $k=p$, where $\phi_{pA}(k)$ is the Fourier transform of an F-P eigenmode $\phi_{pA}(x)$
with a low-lying eigenvalue $\l_{pA}$. 

     Since the FP spectra at the Gribov horizon have been derived here from ordinary 2nd order perturbation theory (plus an ansatz for the
gluon propagator), there is obviously a question of whether perturbation theory can be trusted in this context.   In $D=3$ spacetime dimensions the coupling $g^2$ has units of mass, so the expansion parameter at $p \ra 0$ will be $g^2/m$, while the expansion parameter at large $p$ will
be $g^2/|p|$.  The perturbative calculation of the FP eigenvalue spectrum at $p \ra 0$ should therefore be trustworthy for large $m/g^2$. 
Unfortunately, we have seen that the critical mass parameter $m_c$ corresponding to the Gribov horizon is actually rather small, in units of $g^2$,
with, e.g., $m_c/g^2=0.223$ in Coulomb gauge, and $m_c/g^2=0.0864$ in Landau gauge in three spacetime dimensions and $\a=1$.   Of course, the perturbative expansion may also involve some numerical factors, and without calculating to higher orders, or estimating the radius of convergence in some way, it is difficult to judge the accuracy of the second-order term in the series.   But there is no particular reason for confidence in the second-order results at $m=m_c$ at the quantitative level.   It was argued however in section \ref{FPE}, on rather general grounds, that it is 
natural to expect either Type I or Type II behavior of the Faddeev-Popov spectrum at the Gribov horizon.  The perturbative calculation, at this stage, simply provides a concrete illustration in support of this rather general qualitative argument.

\acknowledgments{This research was supported in part by the U.S.\ Department of Energy under Grant No.\ DE-FG03-92ER40711. }  
    
\appendix*
\section{}

    In order to derive the inequality \rf{ie0} stated in section \ref{FPC}, we begin with the following
\begin{theorem}
 Let $Q$ be any Hermitian operator with a discrete set of eigenstates $\{|n\rangle\}$, whose corresponding eigenvalue spectrum
$\{q_n\}$ is bounded from below, and ordered such that $q_n \le q_{n+1}$. Here the index $n$ runs from 1 up to the dimension of the
Hilbert space $N_H$ (which need not be finite).  Let $\{|\phi_n\rangle\}$ be any other complete set
of orthonormal states spanning the same Hilbert space as the $\{|n\rangle\}$.  Then, for any $N \le N_H$,  
\beq
      \sum_{n=1}^N \langle \phi_n | Q | \phi_n \rangle \ge \sum_{n=1}^N q_n
\eeq
\end{theorem}
\ni This is a fairly trivial generalization of the inequality underlying the Rayleigh-Ritz variational method (the case $N=1$), and the proof
goes as follows:  Define
\bea
       T &\equiv& \sum_{n=1}^N \langle \phi_n | Q | \phi_n \rangle
\non \\
&=&  \sum_{n=1}^N \sum_k \sum_m \langle \phi_n | k \rangle \langle k | Q | m \rangle \langle m |\phi_n \rangle
\non \\
&=& \sum_m q_m P_N(m)
\label{T}
\eea
where 
\beq
        P_N(m) = \sum_{n=1}^N \langle m|\phi_n \rangle \langle \phi_n | m\rangle 
\eeq
Observe that
\beq
        0 \le P_N(m) \le P_{N_H}(m) = 1
\eeq
and
\beq
       \sum_{m=1}^{N_H} P_N(m) = N
\label{constraint}
\eeq
Since $P_N(m) \le 1$, and with
regard to the constraint \rf{constraint}, the smallest possible value of $T$ is 
obviously obtained for
\beq
         P_N(m) = \left\{ \begin{array}{cl}
                            1 & m \le N \cr
                            0 & m > N
                          \end{array}  \right.
\eeq
Substituting this optimal choice into the last line of \rf{T}, we find that  
\beq
      T \ge T_{min} = \sum_{m=1}^N q_m
\eeq
and the inequality stated in the theorem is established.

     From this theorem it follows that
\beq
\sum_{n}^{(N)} (\phi_n|-\nabla^2|\phi_n) \ge \sum_{n}^{(N)} \l^{(0)}_n
\label{ie1}
\eeq
where now the $\{\phi_n \}$ are the eigenstates of the F-P operator, and where we have defined
\beq
           \sum_{n}^{(N)} \equiv \sum_{\bn,A} \theta(N^2 - \bn \cdot \bn)
\label{Nsum}
\eeq
Now let
\beq
\r_{N}(\l) = {1\over N_{colors} V} \sum_{n}^{(N)} \d(\l - \l_n)
\eeq
Assuming non-degeneracy, each $\l$ determines $\bn,A$ uniquely, $\bn$ determines $\l^{(0)}_n$, and $p_n^2 = \l^{(0)}_n$ in the large-volume limit.
In this limit we can therefore we can express \rf{ie1} as
\beq
\int d\l ~ \r_{N}(\l)  (\phi_\l|-\nabla^2|\phi_\l)  \ge \int d\l ~ \r_{N}(\l) p^2(\l)
\label{ie2}
\eeq
At small $\l$ there is some leading power behavior
\bea
 \Bigl\langle \r_{N}(\l)  (\phi_\l|-\nabla^2|\phi_\l) \Bigr\rangle &\approx& a \l^r
\non \\
    \Bigl\langle \r_{N}(\l) p^2_\l \Bigr\rangle &\approx& b \l^q
\eea
Then, since \rf{ie2} must  be true for any mode cutoff $N$, no matter how small, it follows that either $r<q$, or $r=q$ and $a>b$.  Either way,
\beq
\lim_{\l \ra 0}\left\langle {\r_{N}(\l)  (\phi_\l|-\nabla^2|\phi_\l)  \over \l^{1-\e}} \right\rangle \ge   
       \lim_{\l \ra 0} \left\langle {\r_{N}(\l) p^2(\l) \over \l^{1-\e}} \right\rangle
\label{ie3}
\eeq
Finally, given that all the near-zero modes are included in the sum \rf{Nsum}, we have
\beq
\r_{N}(\l) = \r(\l) ~~~\mbox{as}~~~ \l \ra 0
\eeq
and the inequality \rf{ie3} becomes
\beq
\lim_{\l \ra 0}\left\langle {\r(\l)  (\phi_\l|-\nabla^2|\phi_\l)  \over \l^{1-\e}} \right\rangle \ge   
      \lim_{\l \ra 0}  \left\langle {\r(\l) p^2(\l) \over \l^{1-\e}} \right\rangle
\label{ie4}
\eeq
This establishes eq.\ \rf{ie0}.

\end{document}